\documentstyle[emulateapj,flushrt,epsf]{article}

\def \th{\thinspace}

\def \Teff{{$T_{\rm ef\!f} $}}
\def \Mo{{$M_\odot $}}

\def \K{{\th K}}
\def \at{{\rm\char'100}}
\def \apriori{{\it a priori\ }}
\def \eg{{{\it e.g.},\ }}
\def \etal{{\it et al.\ }}

\def \cf{{\it cf.\ }}

\def \ie{{{\it i.e.},\ }}

\def \viz{{\it viz.\ }}

\def \dotd{\hbox{$.\!\!^{\rm d}$}}

\def\SA{{\bf A}}
\def\SB{{\bf B}}
\def\SC{{\bf C}}
\def\SD{{\bf D}}
\def\SE{{\bf E}}

\begin{document}

\submitted{ASTROPHYSICAL JOURNAL, submitted}

\title{Hydrodynamical Survey of First Overtone Cepheids}
\author{Michael Feuchtinger$^{1}$,
        J. Robert Buchler$^{1}$ \&
        Zolt\'an Koll\'ath$^{2}$}

\begin{abstract}

A hydrodynamical survey of the pulsational properties of first overtone
Galactic Cepheids is presented.  The goal of this study is to reproduce their
observed light- and radial velocity curves.  The comparison between the models
and the observations is made in a quantitative manner on the level of the
Fourier coefficients.  Purely radiative models fail to reproduce the observed
features, but convective models give good agreement.

It is found that the sharp features in the Fourier coefficients are indeed
caused by the P$_1$/P$_4 = 2$ resonance, despite the very large damping of the
4th overtone.  For the adopted mass-luminosity relation the resonance center
lies near a period of 4\dotd 2 $\pm$ 0.3 as indicated by the observed radial
velocity data, rather than near 3\dotd 2 as the light-curves suggest.

\end{abstract}


\date{\today}


 {\bigskip
        {\footnotesize
 \noindent $^1$Physics Department, University of Florida, Gainesville, FL, USA;
 buchler\at physics.ufl.edu \\
 \noindent $^2$Konkoly Observatory, Budapest, HUNGARY; kollath\at konkoly.hu
 }}
\section{Introduction}

Historically, s Cepheids denote a certain type of low amplitude Cepheids with
almost sinusoidal light-curves.  Recently, the large microlensing surveys EROS
(Beaulieu \etal 1995), MACHO (Welch \etal 1995) and OGLE (Udalski \etal 1997)
have confirmed unequivocally that these stars are overtone Cepheids.  The vast
majority are first overtone pulsators that coexist with a few second overtone
Cepheids at the lower period end.

  \centerline{\vbox{\epsfxsize=9.0cm\epsfbox{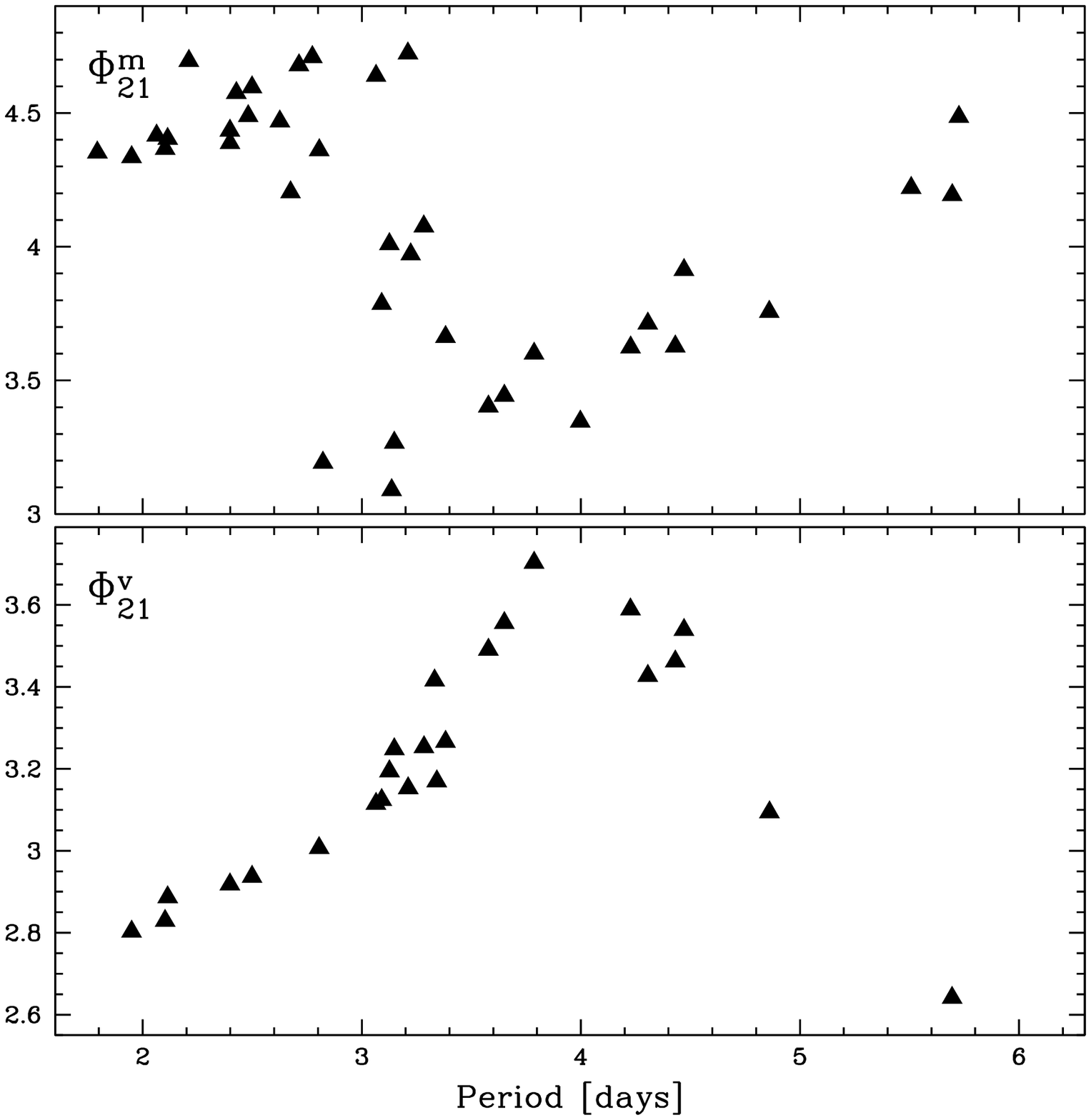}}}

  \noindent{\small Fig.~1: Observational phase difference $\Phi_{21}$ of
      Galactic first overtone Cepheids: light-curves [mag] (upper panel) and
      radial velocity curves (lower panel).}

  \vspace{1cm}

For a comparison between the observational data and the calculated model
pulsations a Fourier decomposition provides an accurate quantitative
representation.  A salient feature in the Galactic first overtone Cepheid
light-curve data (labelled with a superscript m) is a large and sharp drop of
the Fourier phase difference $\Phi_{21}^m$ as a function of period in the
vicinity of the 3\dotd 2 period.  The upper panel of Fig.~1 shows the
observational data summarized in Poretti (1994), and supplemented with V351 Cep
and Anon C Mon (Moskalik, priv. comm.).  The additional Fourier data are
displayed as solid triangles in the left panel of Fig.~3.  The quantity
$\Phi_{31}^m$ exhibits a more or less monotonic, but large $\approx 2\pi$
rise.  The amplitude ratios $R_{21}^m$ and $R_{31}^m$ display a local minimum
in the same vicinity.

A large set of Galactic Cepheid radial velocity data has recently become
available (Kienzle \etal 1999, Krzyt \etal 2000).  The phase difference
$\Phi_{21}^v$ for the radial velocity (superscript v) is plotted in the lower
panel of Fig.~1, and the other Fourier coefficients are displayed as solid
triangles in the left column of Fig.~3.

Rapid variations in the Fourier phases are not special to the first overtone
Cepheids.  Actually, one of the striking features of the classical (fundamental
mode) Cepheids is a Hertzsprung progression of these phases, so named after the
concomitant bump progression that \cite{Hertzsprung} noticed in the shape of
the light-curves.  For the classical Cepheids the center of this progression
lies in the vicinity of the 10 day period.  It was conjectured by Simon \&
Schmidt (1976) that this progression might have its origin in the presence of a
P$_0$/P$_2 = 2$ resonance between the fundamental mode of oscillation and the
second overtone.  This conjecture was later put on a solid mathematical basis
with the help of the amplitude equation formalism (Buchler \& Goupil 1984,
Buchler \& Kov\'acs 1986, Kov\'acs \& Buchler 1989) and was confirmed with
concomitant numerical hydrodynamical modelling (Buchler, Moskalik \& Kov\'acs
1990, Moskalik, Buchler \& Marom 1992).

In fact, it is now well established mathematically that sharp features in the
Fourier coefficients, such as those observed in Cepheids and BL Herculis stars,
are due to the appearance of resonances of the excited mode with an overtone at
certain pulsation periods (\eg \cite{Mito}, Buchler 2000).  Conversely the lack
of such structure as in RR Lyrae is indicative of the absence of resonances.

Subsequently, Antonello \& Poretti (1986), from the behavior of $\Phi_{21}^m$
and $R_{21}^m$ with period (Fig.~3) and from the analogy with the Fourier data
of the fundamental Cepheids, have suggested that a similar resonance, \viz
P$_1$/P$_4 = 2$ is operative in the first overtone Cepheids and is located near
P$_1$ = 3\dotd 2.  However, \cite{Kienzle}, on the basis of the corresponding
radial velocity data, suggest that the resonance center lies at a much higher
period, closer to 4\dotd 6.  This incongruity suggests that it is dangerous to
guess the location of a resonance without proper theoretical input.  We will
discuss this point further below.

In contrast to the fundamental Cepheids, the first overtone Cepheid pulsators
have only received scant theoretical attention.  Aikawa \etal (1987) computed
the radial velocity and light curves of 11 radiative overtone pulsator models
with the specific purpose of reproducing the observations of SU Cas, but were
not satisfied with their results. (One labels radiative models in which
convective heat transport is disregarded).  Later, Antonello \& Aikawa (1993)
calculated two short sequences of Cepheids in order to see if numerical
hydrodynamic modelling would confirm the postulated role of the P$_1$/P$_4 = 2$
resonance in the vicinity of 3 days.  Their results displayed some structure
near the resonance, but failed to reproduce the observed structure in the
Fourier coefficients, in particular the $\Phi_{21}^m$ variation.  The number of
computed models was rather limited, and artificially enhanced opacities were
used.  Subsequently, Schaller \& Buchler (1994), on the basis of an extensive
study of radiative first overtone Cepheids with the OPAL opacities, reached the
conclusion that radiative models cannot reproduce the observed structure of the
Fourier coefficient $\Phi_{21}^m$; a similar conclusion was also reached by
Antonello \& Aikawa (1995).  This disagreement with observation came as a
surprise considering how well the fundamental mode Galactic Cepheid pulsations
can be modelled (\eg Moskalik, Buchler \& Marom 1992).

In the last few years a lot of effort has been devoted to including convection
in the pulsation codes, and recently one of the major remaining challenges,
namely the modelling of beat pulsations, has been met (\eg Koll\'ath \etal
1998, Feuchtinger 1998).  In this paper we apply the same convective codes to
the study of first overtone pulsations.

\section{Physical input}

Linear and nonlinear models are calculated with the Vienna pulsation code
(Feuchtinger 1999a), which solves the equations of radiation hydrodynamics
together with a time-dependent model equation for turbulent convection.  This
code recently has been extended by a linear nonadiabatic normal mode analysis,
details of which are presented in a separate paper.  For comparison purposes
some of the calculations that are described in this paper have also been
performed in parallel with the Florida pulsation code (described in Koll\'ath,
Buchler, Szab\'o \& Csubry 2000).  The latter uses a different numerical
approach, but with only minor differences in the input physics.  We have
ascertained that the two codes give basically the same results.

For the Rosseland mean of the opacity we use the most recent OPAL tables
(Iglesias \& Rogers 1996) which are augmented by the Alexander \& Ferguson
(1994) low temperature opacities below 6000\K.  The Eddington factor is set to
1/3.

  \centerline{\vbox{\epsfxsize=9.0cm\epsfbox{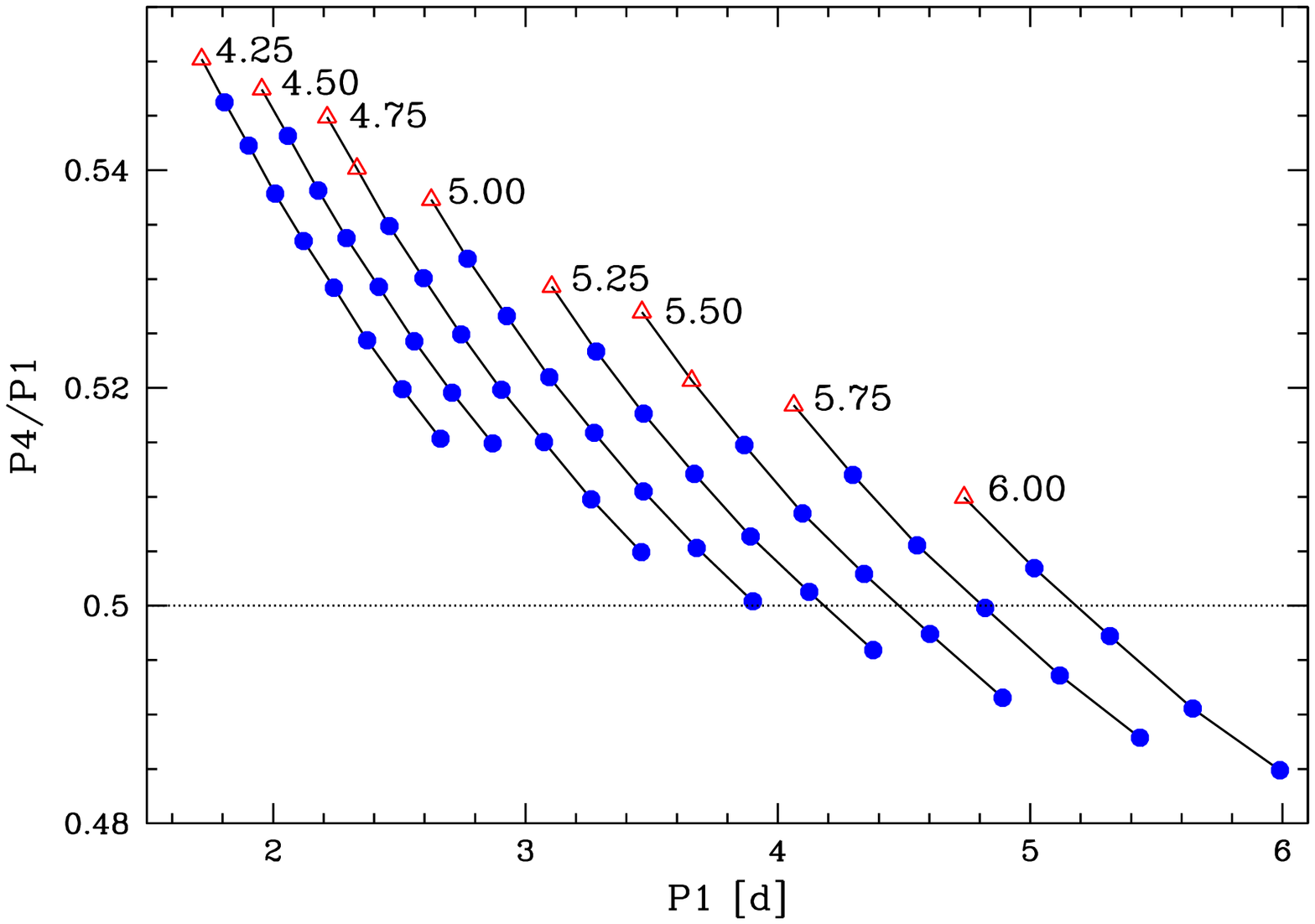}}}

  \noindent{\small Fig.~2: Period ratio $P_4/P_1$ versus pulsation period for
           radiative models. Open triangles refer to vibrationally
           stable models and filled circles to models which have a stable
           limit-cycle. Labels on top indicate the stellar mass of
           the corresponding sequence.}

  \vspace{1cm}

Our {\it model sequences} have constant mass and luminosity and an equilibrium
effective temperature varying in steps of 100\K.  They represent horizontal
paths through the instability strip (IS).  We adopt a mass--luminosity (ML)
relation: \ \ $\log(L/L_{\odot}) = 0.79 + 3.56 \log(M/M_{\odot})$,\ \ which is
derived from the stellar evolution calculations of Schaller \etal (1992) which
make use of the same OPAL opacity data.  For a good coverage of the observed
period range we vary the stellar mass between 4.25 and 6.5 \Mo\ in steps of
0.25 \Mo. The chemical composition corresponds to a typical Galactic one of
(X,Y,Z) = (0.70, 0.28, 0.02).

\begin{figure*}
  \vspace{0cm}
  \centerline{\vbox{\epsfxsize=18cm\epsfbox{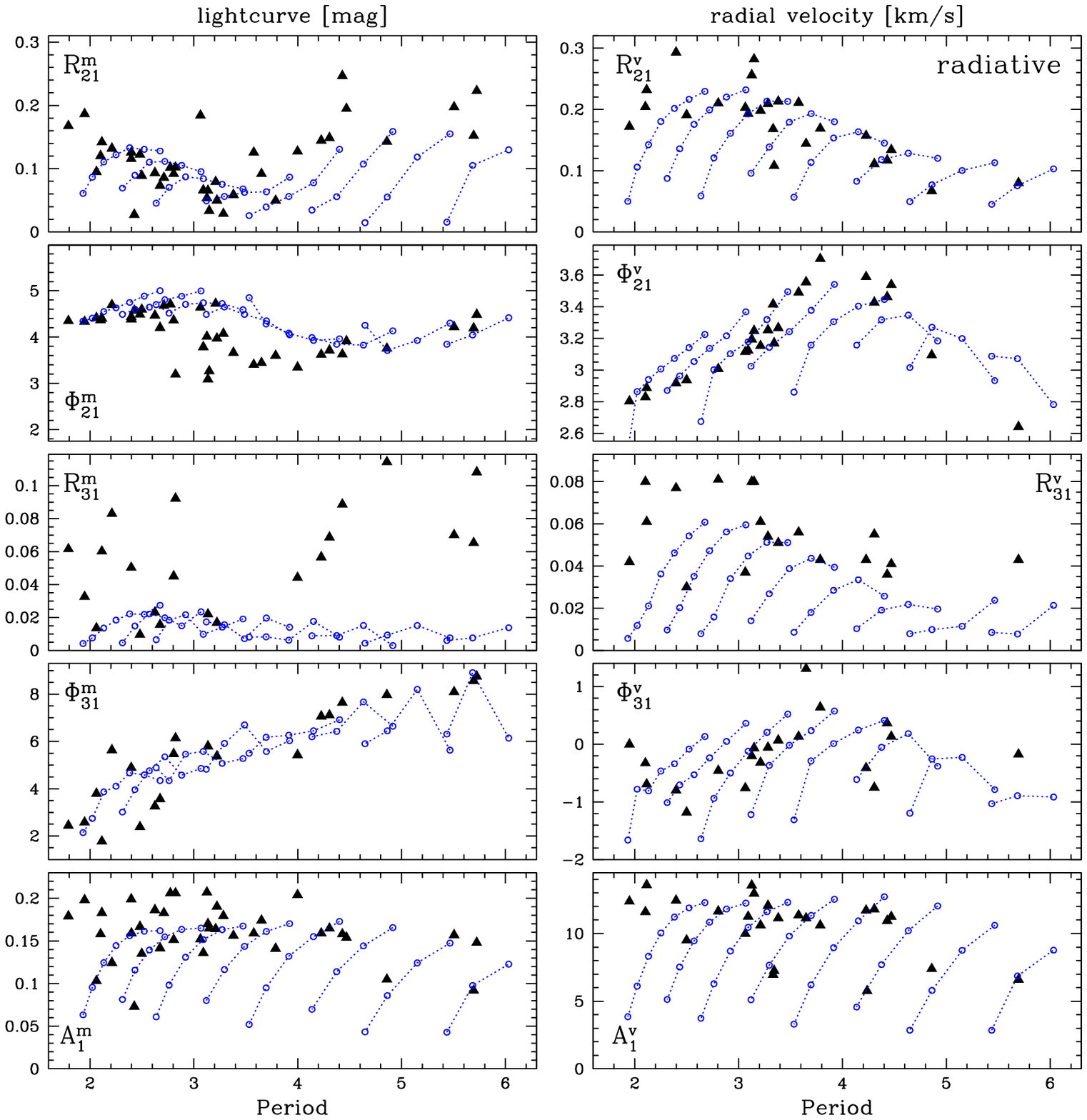}}}

  \noindent{\small Fig.~3: Fourier coefficients of radiative models (open
           circles) compared to observations (filled triangles): {\sl Left:}
           light-curve (mag) data; {\sl Right:} radial velocity (km/s) data.
           The open circles connected by dotted lines refer to sequences with
           the same mass, starting from M = 4.25 \Mo\ at the left to M = 6.25
           \Mo\ at the right in steps of 0.25 \Mo.}

\end{figure*}

The Fourier decomposition of the resulting light- and radial velocity curves is
calculated by a least squares fit with a standard Fourier sum (8 terms).
Amplitude ratios $R_{n1} = A_n/A_1$ and phase differences $\Phi_{n1} = \Phi_n -
n \Phi_{1}$ are then used for the comparison to the observed data.  Following
custom, a cos Fourier decomposition is used for the light-curve data, and a sin
decomposition for the radial velocity data.  Note further that we compute
bolometric light variations, which are compared to V-band magnitudes.  For the
case of RR Lyrae stars it has been shown that the differences in the low order
Fourier coefficients between bolometric and V light-curves are rather small, in
particular for low amplitude first overtone pulsations (Dorfi \& Feuchtinger
1999, Feuchtinger \& Dorfi 2000).  However, for metal-rich Galactic Cepheids
this has to be checked by detailed radiative transfer calculations, which will
be done in a companion paper.

For the transformation between theoretical and observed radial velocities we
apply a constant projection and limb darkening factor ($u_{\rm obs} = u_{\rm
cal} / 1.4$) to the calculated velocity values (Cox 1980).

\section{Radiative models}

As a first step we reexamine the difficulties encountered by radiative
pulsation models, \ie models that for simplicity disregard all convection.

  \vskip 10pt
  \centerline{\vbox{\epsfxsize=9.0cm\epsfbox{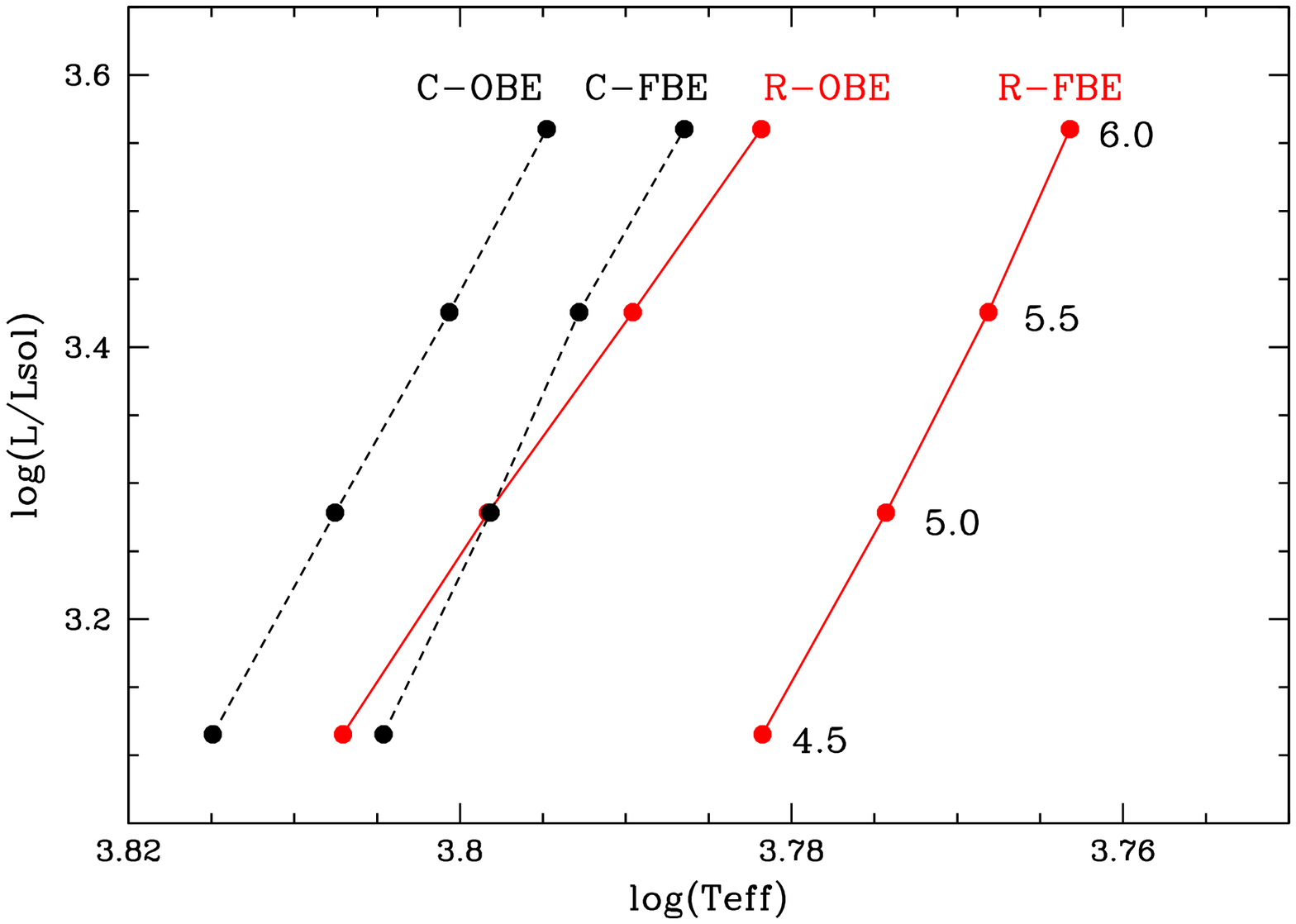}}}

  \noindent{\small Fig.~4: Linear blue edges of {\it radiative} models (R-FBE
           and R-OBE) compared with those of convective models, series \SA\
           (C-FBE and C-OBE).  The labels on the right indicate the stellar
           masses.}

  \vspace{1cm}

As already discussed in the Introduction, a resonance between the first and the
fourth overtone is responsible for the characteristic variations in the Fourier
coefficients of both light- and radial velocity variations.  The location of this
resonance with respect to the pulsation period, which is of particular
importance for the interpretation of nonlinear results, can best be determined
from linear results.  In Fig.~2 the period ratio $P_4/P_1$ is plotted as a
function of $P_1$ for each sequence of constant mass, and filled circles denote
models with a stable overtone limit-cycle.  Models close to the resonance
center ($P_1/P_4 = 2$), which fall within the region of stable overtone
pulsation, appear between about P$_1$ = 4 and 5\dotd 2.

The results of the nonlinear radiative survey are summarized in Fig.~3 which
depicts the low order Fourier coefficients, on the left for the light-curves,
and on the right for the radial velocity curves.  The observational data are
represented by filled triangles, the theoretical models by open circles with
dotted lines connecting the models of each sequence.  We recall that the
sequences consist of models with a given mass and luminosity, with \Teff\
decreasing and $P_1$ increasing to the right.

Even though the overall picture is not at all disastrous, several severe
problems are visible.  First, and most strikingly, from the flatly distributed
theoretical $\Phi_{21}^m$ it is evident that the Z-shape of the observed data
cannot be reproduced at all -- a disagreement which has already been mentioned
in the Introduction.  In addition, the theoretical $R_{21}^m$ values are too
low for periods greater than 4 days, and the $R_{31}^m$ are much too low
overall.  In contrast, the $\Phi_{31}^m$ show reasonable agreement.

While the overall level of the pulsation amplitudes is set by pseudo-viscosity,
it is interesting that the behavior of the amplitudes $A_1^m$ and $A_1^v$ as a
function of P$_1$ follows the observations rather well.

For the radial velocity plots, the general agreement with observations is much
better than for the light-curves.  In particular, the calculated data fit the
observed $\Phi_{21}^v$ distribution.  However, several models lie off the well
defined observational distribution.  The same discrepancy is also visible in
all the other quantities.  Below we show that the inclusion of convection
gives better agreement.

In summary we thus corroborate the fact that radiative models are not able to
reproduce satisfactorily the observational behavior of first overtone Cepheid
pulsations.

\section{Convective models} \label{s:conv}

In the last few years it has become evident that the inclusion of convective
energy transport is critical to the modelling of classical stellar pulsations,
rather than just being necessary for stabilizing the models at low \Teff.  The
unpleasant consequence is that several free parameters ($\alpha$'s) have to be
added to the former parameter-free radiative pulsation models.  Theory
unfortunately provides no guidance for choosing the values of these parameters,
and therefore a calibration with observational data becomes necessary (\eg
Stellingwerf 1984, Yecko \etal 1998, Feuchtinger 1999a).

  \vskip 10pt
  \centerline{\vbox{\epsfxsize=9.9cm\epsfbox{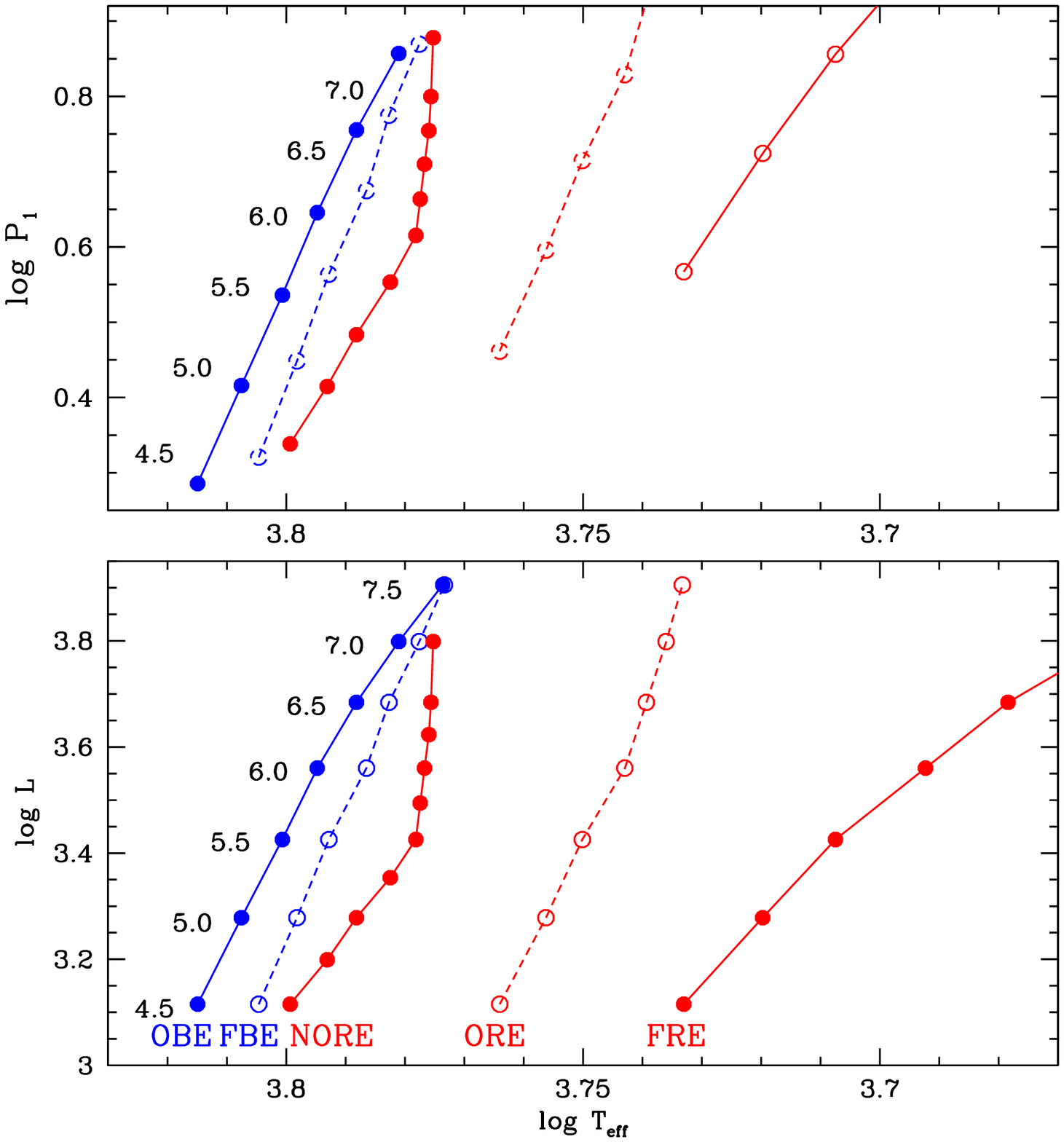}}}
  \noindent

{\small Fig.~5: Instability strip boundaries for {\it convective} models
           (series \SA), bottom:  in the Log L - Log \Teff\ plane,
           top: in the Log P$_1$ - Log \Teff\ plane.  From left to right:
           (first) overtone linear blue edge (OBE), 
           fundamental linear blue edge (FBE), 
           nonlinear overtone red edge (NORE), 
           overtone linear red edge
           (ORE) and fundamental linear red edge (FRE);
           the labels on left refer to the stellar masses.}

  \vspace{1cm}

For the present investigation we use the convection model according to
Kuhfu\ss~(1986) and Gehmeyr \& Winkler (1992) in the version of Wuchterl \&
Feuchtinger (1999).  Essentially the same model has been adopted by the Florida
pulsation code (\cf Koll\'ath \etal~2000), but with a slightly different
parameterization.  A summary of the free parameters (subsequently termed
$\alpha$'s) and the interrelations between the two sets of parameters are given
in Table~1.  For details we refer to the above cited references.

\begin{table*}
  \vspace{0.3cm}
 {\small

  \noindent{Table~1: Free parameters of the time-dependent turbulent convection
           model.  Columns 2 and 3 list the free parameters as defined in the
           Vienna and the Florida codes, respectively.  Column 4 (interrelation)
           gives the Florida values as a function of the Vienna values.  Columns
           \SA\ through \SE\ give the adopted parameter sets for our five model
           series in terms of the Vienna parameterization.  No interrelation is
           given for the radiative cooling, as this effect is modelled slightly
           differently in the two codes. The parameters $\bar\alpha_{s}$,
           $\bar\alpha_{c}$ and $\bar c_{D}$ are normalized to their
           standard values as given in the text. }

  \vspace{0.3cm}
  \begin{center}
  \begin{tabular}{lcclllllll}
    \hline
    \hline
    \noalign{\smallskip}
    Physical meaning  &  Vienna code & Florida code &  interrelation &
    Series: & \SA & \SB & \SC & \SD & \SE \\
    \noalign{\smallskip}
    \hline
    \hline
    \noalign{\smallskip}
    \noalign{\smallskip}
    mixing length         & $\alpha_{\scriptscriptstyle \rm ML}$  & $\alpha_{\Lambda}$
                          & $\alpha_{\scriptscriptstyle \rm ML}$
                          & & 1.5 & 1.5 & 2.0 & 1.5 & 1.5 \\

    \vspace{0.05cm}
    turbulent source      & $\bar\alpha_{s}$ & $\bar\alpha_{s}$
                          & $\sqrt{\bar\alpha_{s}/\bar c_D}$
                          & & 1 & 1 & 1 & 1 & 1 \\

    \vspace{0.1cm}
    turbulent dissipation &   $\bar c_D$ & $\bar\alpha_{d}$
                          &   $\bar c_D$
                          & & 1 & 1 & 1 & 4 & 4 \\
    \vspace{0.1cm}
    convective flux       & $\bar\alpha_{c}$ & $\bar\alpha_{c}$
                          & $\bar\alpha_{c} \bar\alpha_{s}$
                          & & 1 & 1 & 1 & 1.5 & 1.5 \\
    \vspace{0.05cm}
    overshooting          & $\alpha_{t}$ & $\alpha_{t}$
                          & $\alpha_{t} / c_D$
                          & & 0 & 0 & 0 & 0 & 0.001 \\
    \vspace{0.05cm}

    turbulent viscosity   & $\alpha_{\mu}$ & $\alpha_{\nu}$
                          & $\alpha_{\mu}$
                          & & 0.25 & 0.33 & 0.35 & 0.50 & 0.50\\

    \vspace{0.1cm}
    turbulent pressure    & $\alpha_{p}$ & $\alpha_{p}$
                          & $\alpha_{p}$
                          & & 0 & 0 & 0 & 0 & 2/3 \\

    \vspace{0.1cm}
    flux limiter          & $\alpha_{L}$
                          & $Y_{\rm lim}$ & --
                          & & 0 & 3 & 0 & 0 & 0 \\
    \vspace{0.1cm}
    radiative cooling     & $\gamma_{\rm R}$ & $\alpha_{R}$
                          & --
                          & & 0 & 0 & 3.5 & 0 & 0 \\

    \hline
    \hline
  \end{tabular}
  \end{center}
 }
\end{table*}

In the following we present five series of calculations, \SA\ through \SE,
whose $\alpha$'s are given in Table~1.  In order to reduce the multidimensional
parameter space to a reasonable set of $\alpha$'s, we have pursued the
following strategy.  The parameters $\alpha_{s}$, $c_D$ and $\alpha_{c}$ can be
chosen to reduce the model to mixing length theory in the local static limit
(Kuhfu\ss~1986, Wuchterl \& Feuchtinger 1998), for the values $\alpha_{s} = 1/2
\sqrt{2/3}$, $c_D = 8/3 \sqrt{2/3}$ and $\alpha_{c} = \alpha_{s}$.  The
quantities $\bar\alpha_{s}$, $\bar\alpha_{c}$ and $\bar c_{D}$ in Table~1 are
given relative to these 'standard' values.  We adopt the standard values in
series \SA, and in addition set the mixing length parameter
$\alpha_{\scriptscriptstyle \rm ML}$ to the widely used value of 3/2.  The
parameter of the turbulent viscosity $\alpha_{\mu}$ is used to adjust the
pulsation amplitude.  Turbulent pressure, overshooting, radiative losses and
the convective flux limiter are disregarded in \SA.  Series \SB\ investigates
the effects of the flux limiter and series \SC\ the effects of radiative
losses.  Series \SD\ has much lower turbulent energy than series \SA\, and
series \SE\ additionally includes the turbulent pressure and the turbulent
flux.  We wish to emphasize that the adopted choices of free parameters are by
no means unique.

\vskip 10pt
\centerline{Instability Strip}
\vskip 10pt

First we examine the influence of convection on the blue edge for series \SA\
and compare it to the radiative models.  In Fig.~4 the radiative linear blue
edges (R-FBE and R-OBE) are drawn as solid lines, and the convective ones
(C-FBE and C-OBE) as dotted lines.  In contrast to the frequently adopted
notion that convection is only important near the red edge of the IS (\cf
however Stellingwerf 1984), both the fundamental and the first overtone blue
edges are shifted toward higher temperatures (toward the left in the figure)
by about 350\K\ and 150\K\ for the fundamental and first overtone pulsations,
respectively.

\begin{figure*}
  \centerline{\vbox{\epsfxsize=18cm\epsfbox{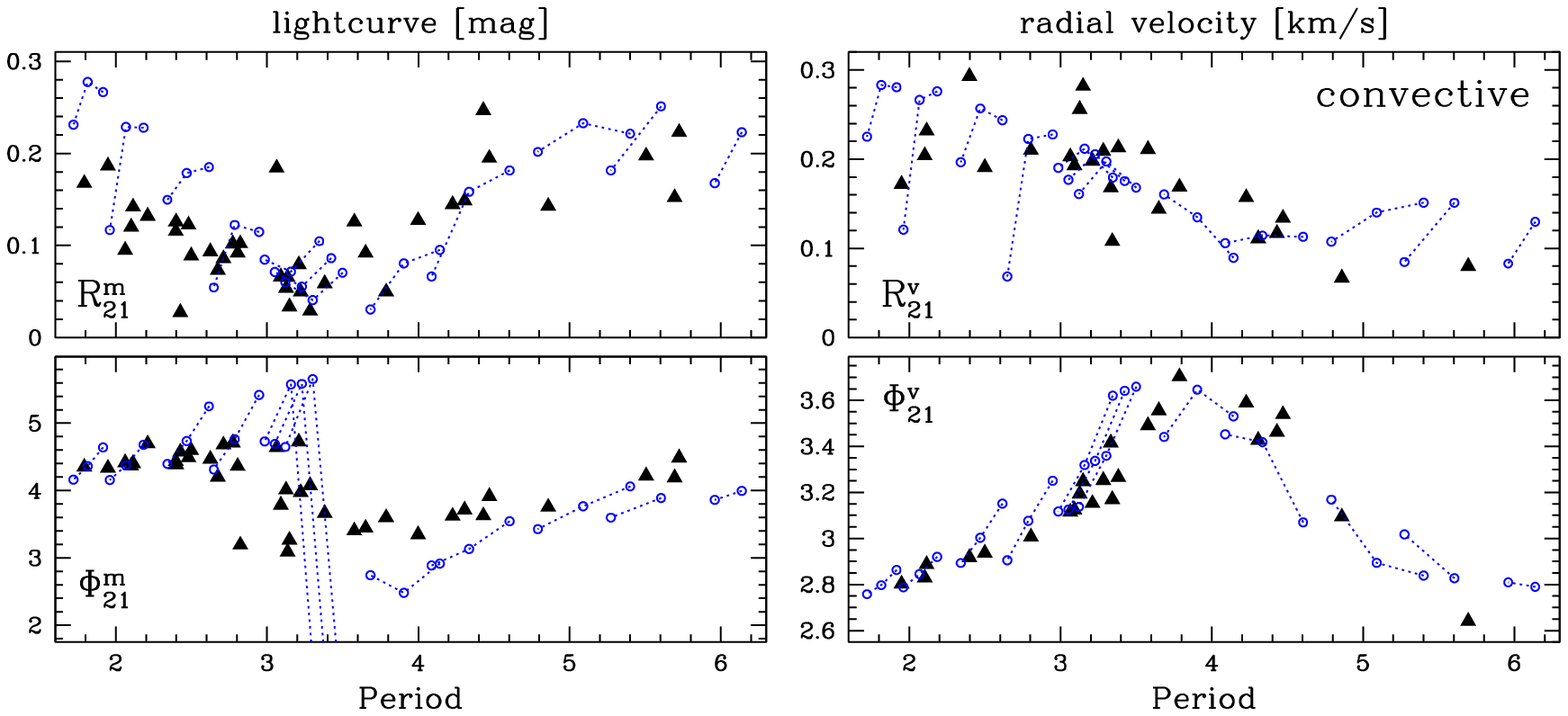}}}

  \noindent{\small Fig.~6: Fourier coefficients of convective models, series
           \SA\ (open circles) compared to observations (filled triangles):
           {\sl Left:} light-curve (mag) data; {\sl Right:} Radial velocity
           (km/s) data.  The open circles connected by dotted lines refer to
           sequences with the same mass (from left to right: 4.25, 4.50, 4.75,
           5.00, 5.15, 5.20, 5.25, 5.50, 5.75, 6.0, 6.25, 6.50\Mo).}
\end{figure*}

The complete {\it linear} topography of the IS for series \SA\ is presented in
Fig.~5.  The dotted lines refer to fundamental mode pulsation and the solid
lines to the first overtone, filled/open circles to blue edge/red edges.  These
edges are somewhat sensitive to the values of the $\alpha$'s (\eg Yecko \etal
1998) and we show a comparison of the three series below.

The {\it nonlinear} first overtone red edge (NORE) is plotted as a dashed line
in Fig.~5.  It is located at considerably higher temperatures than the
corresponding linear one.  Slight smoothing has been applied because of the
rather coarse steps in effective temperature.  The low mass models (with M $<$
5.5\Mo) that are located at the right side of the NORE are double-mode
pulsators, whereas the more massive ones (M $>$ 5.5\Mo) pulsate in the
fundamental mode.  This modal change is the reason for the kink in the NORE
(\cf Koll\'ath \etal~1998, Koll\'ath \etal~2000) for a detailed picture of the
modal selection problem).

It is important to exercise considerable care that the computed overtone
limit-cycles are indeed stable, and not just on a transient to either
double-mode or to fundamental pulsations.  These transients can be very long
lasting and give an erroneous impression of steady behavior.  A very efficient
way of determining this stability with the 'analytical signal' method is
discussed in Koll\'ath \& Buchler (2000).

As expected and already discussed earlier (Yecko \etal 1998 and \cite{KBBY}),
the fundamental and first overtone blue edges intersect at some point (at
$\sim$ 7.5 \Mo).  This is consistent with the observational fact that the
overtone Cepheid periods exhibit an upper limit, which is around P$_1$ = 6 days
for the Galaxy (with one single star found at 7\dotd 57).  The linear overtone
period at the intersection point is 8\dotd 9 here which is considerably higher
than the observations suggest.  However, the region above 6.5\Mo\ where stable
overtone pulsations are possible is very narrow, which reduces the
observational likelihood of such long period first overtone Cepheids.
Furthermore the linear growth rates are found to be very small, and the
corresponding nonlinear models exhibit tiny amplitudes (around 0.03$^m$ for the
7 \Mo\ sequence), since the pulsation amplitude scales with the square root of
the growth-rate ($A \sim$ $\sqrt \kappa$).  From the nonlinear survey we find
that the maximum overtone period lies close to the observed one only when the
pulsation amplitudes are in general agreement with observed ones.  Our efforts
to adjust the $\alpha$'s so as to lower the period at the intersection point,
reduce the growth-rates and the pulsation amplitudes too much.

  \centerline{\vbox{\epsfxsize=9.6cm\epsfbox{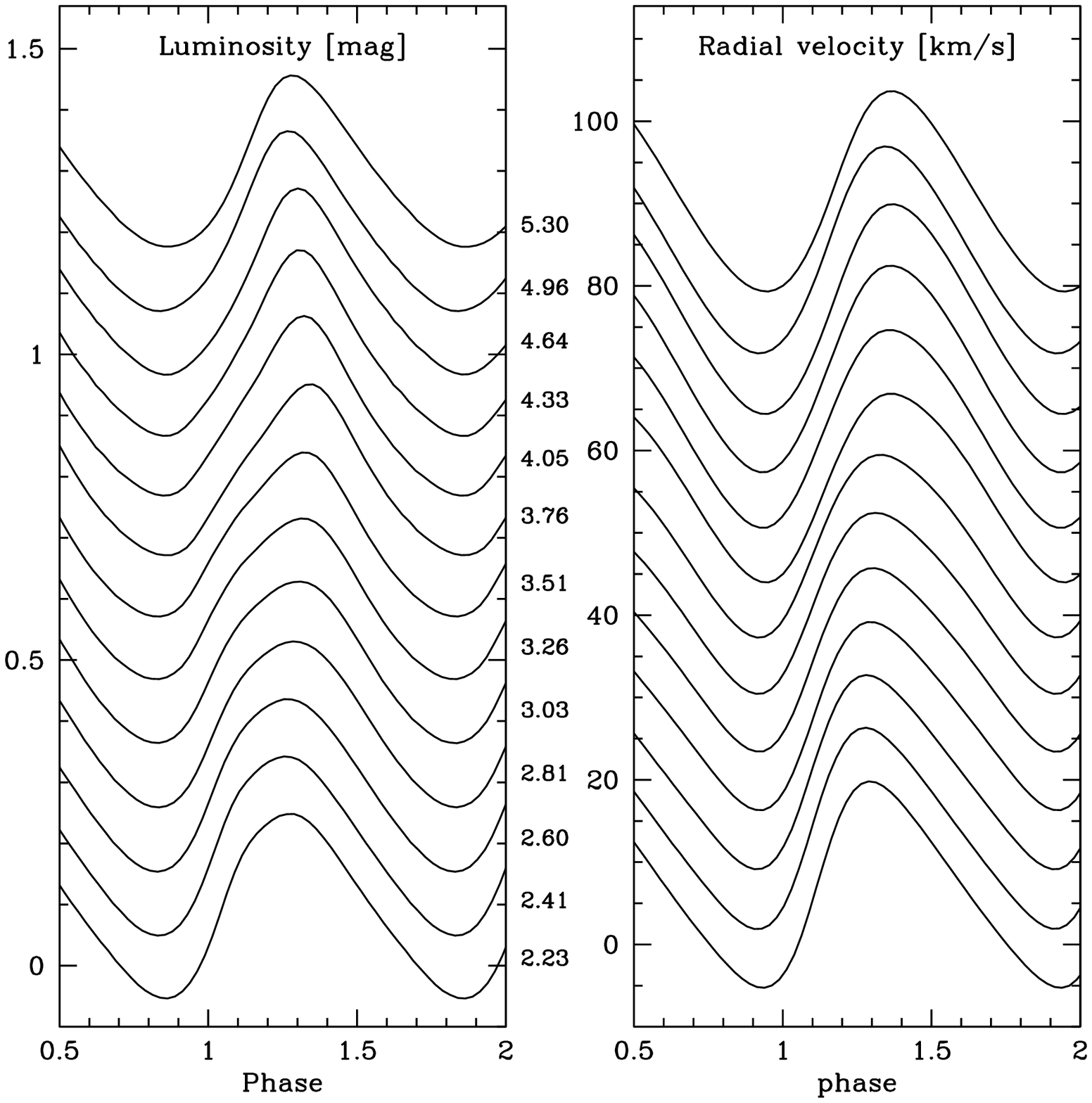}}}

  \noindent{\small Fig.~7: {\sl Left:} Light curves and {\sl right:} 
            radial velocity
            curves for a sequence parallel to the blue edge.
            The light-curves are shifted vertically by 0.1 mag and the radial
            velocity curves by 7 km/s. The curves are labelled with the
            periods.}

 \vspace{1cm}

\vskip 10pt
\centerline{Light-Curves and Radial Velocities}
\vskip 10pt

Fig.~6 displays the light- and radial velocity curve data for series \SA.
Again we use filled triangles for the observations, while open circles
represent our full amplitude pulsating models.

The light-curve Fourier coefficients for series \SA, exhibited on the left of
Fig.~6, show great improvement with respect to the radiative series.  The
theoretical $\Phi_{21}^m$ distribution attracts immediate attention with a very
conspicuous jump around P$_1$ = 3\dotd 4, in contrast to all the radiative
models.  In fact, the last points of the sequences 5 through 7 fall in the
range 0.0--1.0, way below the scale.  Even though the magnitude of the jump is
considerably higher than what is observed, our model series reproduces
qualitatively the observational $\Phi_{21}^m$ behavior.  In addition, all other
light-curve Fourier coefficients show good overall agreement with observations.
The values of the 31 Fourier coefficients are practically the same for series
\SA\ through \SE\ and we refer to Fig.~10 for their display.  Compared to the
radiative models there is an average increase in $R_{31}^m$ by almost a factor
of 10, and for small periods, the convective models also display higher
pulsation amplitudes and $R_{21}^m$ values.

The radial velocity data on the right of Fig.~6 show good overall agreement as
well.  In particular, the $R_{21}^v$ and $\Phi_{21}^v$ distributions closely
follow the observed ones, and they produce a much better match than the
radiative models.  For $R_{31}^v$ and $\Phi_{31}^v$ a similar behavior occurs,
even though the $R_{31}^v$ lie somewhat below the observed ones (\cf Fig.~10).
However, the $R_{31}^v$ are tiny which decreases the relevance of this
deviation.  The only perhaps significant discrepancy appears in the calculated
amplitudes which, for the higher pulsation periods, are larger than the
observed ones.  This is also reflected in the larger $R_{21}^v$.

We have used the observed overall value of the pulsation amplitude to calibrate
the $\alpha$'s (in practice $\alpha_\mu$).  When the amplitudes are increased
beyond the observed values the jump in $\Phi_{21}^m$ becomes increasingly weak
and in disagreement with the observations.  

  \centerline{\vbox{\epsfxsize=9.0cm\epsfbox{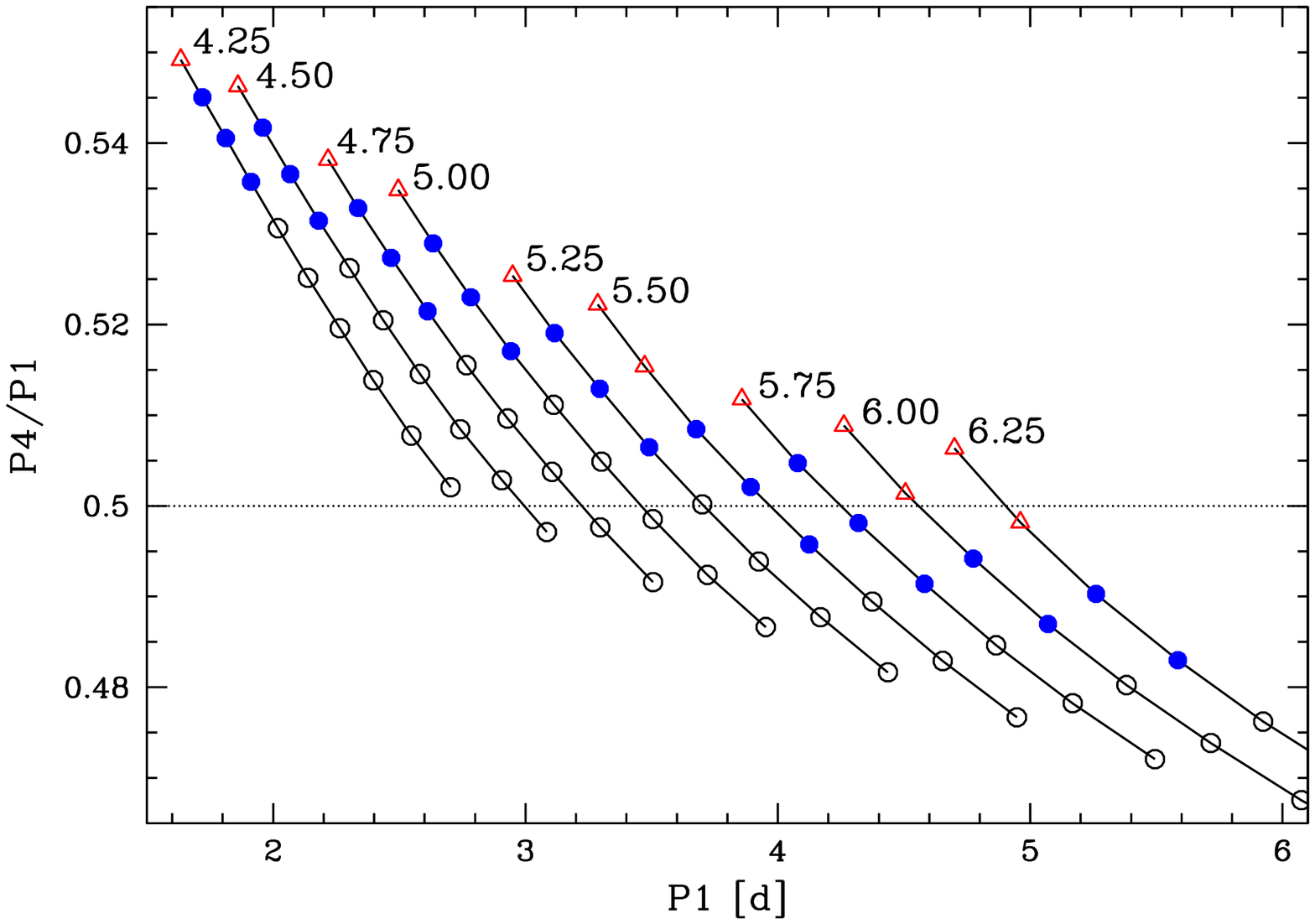}}}

  \noindent{\small Fig.~8: Period ratio $P_4/P_1$ versus pulsation period for
           convective models (series \SA).  Open triangles denote vibrationally
           stable models. Filled/Open circles
           refer to models with a stable/unstable overtone limit-cycle.
           The labels on the right indicate the stellar masses.}

  \vspace{1cm}

The shapes of the calculated light- and radial velocity curves are displayed in
Fig.~7 for a sequence of models running at 200\K\ distance parallel
to the overtone blue edge.

Finally, we note that we have computed the same series \SA\ with the Florida
convective (Lagrangean) pulsation code, and that the results are essentially
identical.  Despite the Lagrangean nature of the latter calculations the models
show a very smooth behavior, in contrast to the radiative models for which the
adaptive code is necessary (Buchler, Koll\'ath \& Marom 1996) to give smooth
light-curves (cf.~also Sect.~\ref{s:lag}).

In summary we emphasize that the inclusion of convection is crucial for a
successful quantitative modelling of the pulsational properties of first
overtone Cepheids, in particular of the Fourier decomposition coefficients of
the light- and radial velocity curves.

\vskip 10pt
\centerline{Location of Resonance}
\vskip 10pt

We return here to the important question of whether the resonance center is
near P$_1$ = 3\dotd 2 as suggested by the light-curves (Antonello \& Poretti
1986) or near 4\dotd 6 as the radial velocity data indicate (Kienzle \etal
1999).

First, we note that our calculations which used the Schaller \etal M--L
relation ($\log(L/L_{\odot}) = 0.79 + 3.56 \log(M/M_{\odot})$), reproduce the
observed shift with period between the light-curve and the radial velocity
curve $R_{21}$ and $\Phi_{21}$.  From our calculated linear period ratios we
should therefore be able to locate the resonance center, and resolve this
issue.  (We stress that it is important to use the same code, \ie the same
differencing scheme and the same mesh to compare the hydrodynamics results to
the linear periods).  We note in passing that the relative differences between
the nonlinear and the linear periods are at most +0.4\%.

The linear period ratios P$_4$/P$_1$ versus pulsation period for our convective
series \SA\ are shown in Fig.~8.  The filled circles denote models with a
stable nonlinear overtone limit-cycle.  Note that our nonlinear first overtone
IS is very narrow.  We shall return later (\S6) to the importance of the
narrowness.  Only two of our mass sequences (5.5 and 5.75 \Mo) can undergo
stable overtone pulsations with periods near the resonance center (in contrast
to the radiative models of Fig.~2).  The corresponding pulsation periods reveal
the resonance to be located around P$_1$ = 4\dotd 2 $\pm 0.3$, in fact very
close to the value of 4\dotd 6 that Kienzle \etal (1999) had conjectured.

Our calculations leave no doubt that the P$_1$/P$_4$ = 2 resonance is
responsible for the observed structure of the light and radial velocity Fourier
coefficients, and that the resonance is located in the vicinity of P$_1$ =
4\dotd 2.
 
It is somewhat surprising that the 2:1 resonance with the fourth overtone has
such a pronounced effect on the Fourier data, because after all this overtone
is so strongly damped.  It has a relative damping rate per pulsation period of
$\kappa_4 $P$_0 \sim -0.4$\th in the vicinity of the resonance, \ie its
amplitude would decay by 33\% in one pulsation period.

  \centerline{\vbox{\epsfxsize=9cm\epsfbox{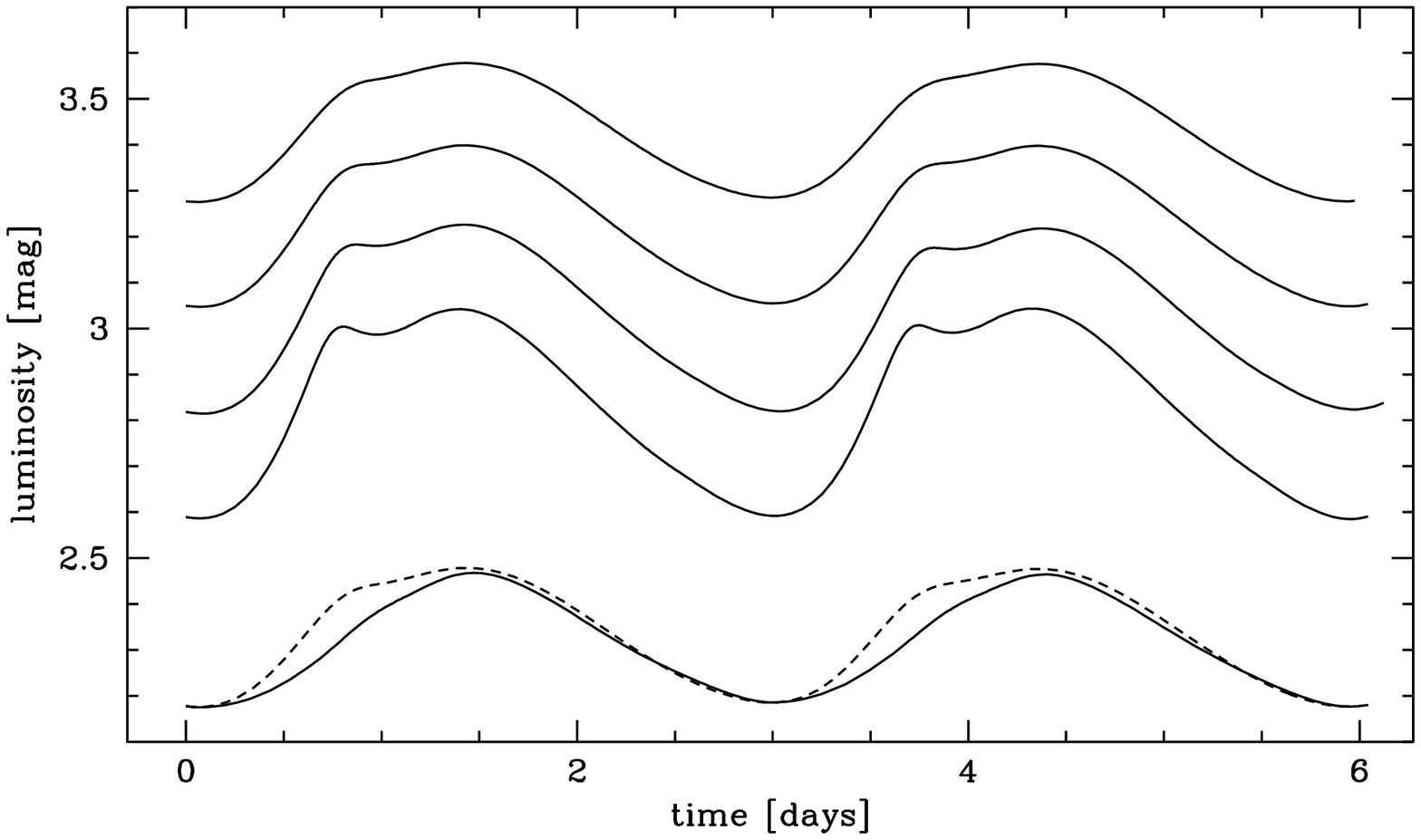}}}

  \noindent{\small Fig.~9: Effect of pulsation amplitude on the light-curve for
            a series \SA\ model located at the $\Phi_{21}^m$ jump.  Upper four
            solid lines, have decreasing turbulent viscosity, 0.25 (top) to 0.1
            (bottom) in steps of 0.05.  The lowest solid line shows the
            corresponding light-curve with the convective flux limiter
            included, the dashed line refers to the same model without the flux
            limiter.}

  \vspace{1cm}

\section{Sensitivity to numerical and physical input} \label{s:sensitivity}


\subsection{Lagrangean versus adaptive mesh} \label{s:lag}

We mentioned in the preceding paragraphs that a comparison between convective
Lagrangean and adaptive calculations reveals no differences, as far as first
overtone Cepheid models are concerned.  This comes as no surprise, because
pulsation amplitudes are rather small, and no strong shock waves appear in the
dynamics which would require a more elaborate numerical treatment.  Moreover,
the inclusion of convective energy transport considerably smoothes the sharp
features in the combined H--He ionization zone which are a well known headache
for radiative modelling.

However, a word of caution is necessary here.  As already discussed in detail
in Feuchtinger \& Dorfi (1994) and Buchler, Koll\'ath \& Marom (1996), adaptive
models suffer from advection errors due to the non-Lagrangean motion of the
cell boundaries.  These errors are particularly severe in the interior where
the cell-masses increase rapidly.  In order to keep these errors small, the
interior part of the model has to be treated as Lagrangean.  The switching
point between Lagrangean and adaptive zoning therefore has to be chosen with
some care, as advection errors can considerably influence the dynamical
behavior and ultimately the morphology of the light- and radial velocity curve.
By comparing the adaptive results to Lagrangean results we checked in
detail that our results are not vitiated by advection errors.

\subsection{Radiation hydrodynamics versus equilibrium diffusion}

A standard radiation diffusion equation for radiative transport is much more
convenient and faster than a time-dependent treatment of radiative transfer
(radiation hydrodynamics).  Since both codes are available, it has seemed
interesting to check whether the simplified diffusion was adequate for
pulsational behavior.  On the basis of the study of several sequences of models
we find that, apart from small changes in the pulsation amplitudes, the
results are essentially the same for both treatments.  In particular no
noticeable effect on the low order Fourier coefficients has been found.
A radiation diffusion treatment is therefore fully adequate.

\subsection{The M--L Relation}

Our results do not depend sensitively on the chosen M--L relation as long as
the latter puts the resonance in the right place.  This is so because the
agreement of the hydrodynamical results with the observations necessarily puts
the resonance at the right place and thus fixes the zero-point of the M--L
relation (Buchler \etal 1996).  The properties of the models depend very little
on the slope of the M--L relation because of the relatively narrow mass range
of the overtone Cepheids.

\subsection{Convection and the $\alpha$ parameters}

In the following we discuss how several of the convective parameters influence
the behavior of first overtone Cepheid models and in particular the Fourier
coefficients of the light- and radial velocity curves.

\begin{figure*}
  \centerline{\vbox{\epsfxsize=18cm\epsfbox{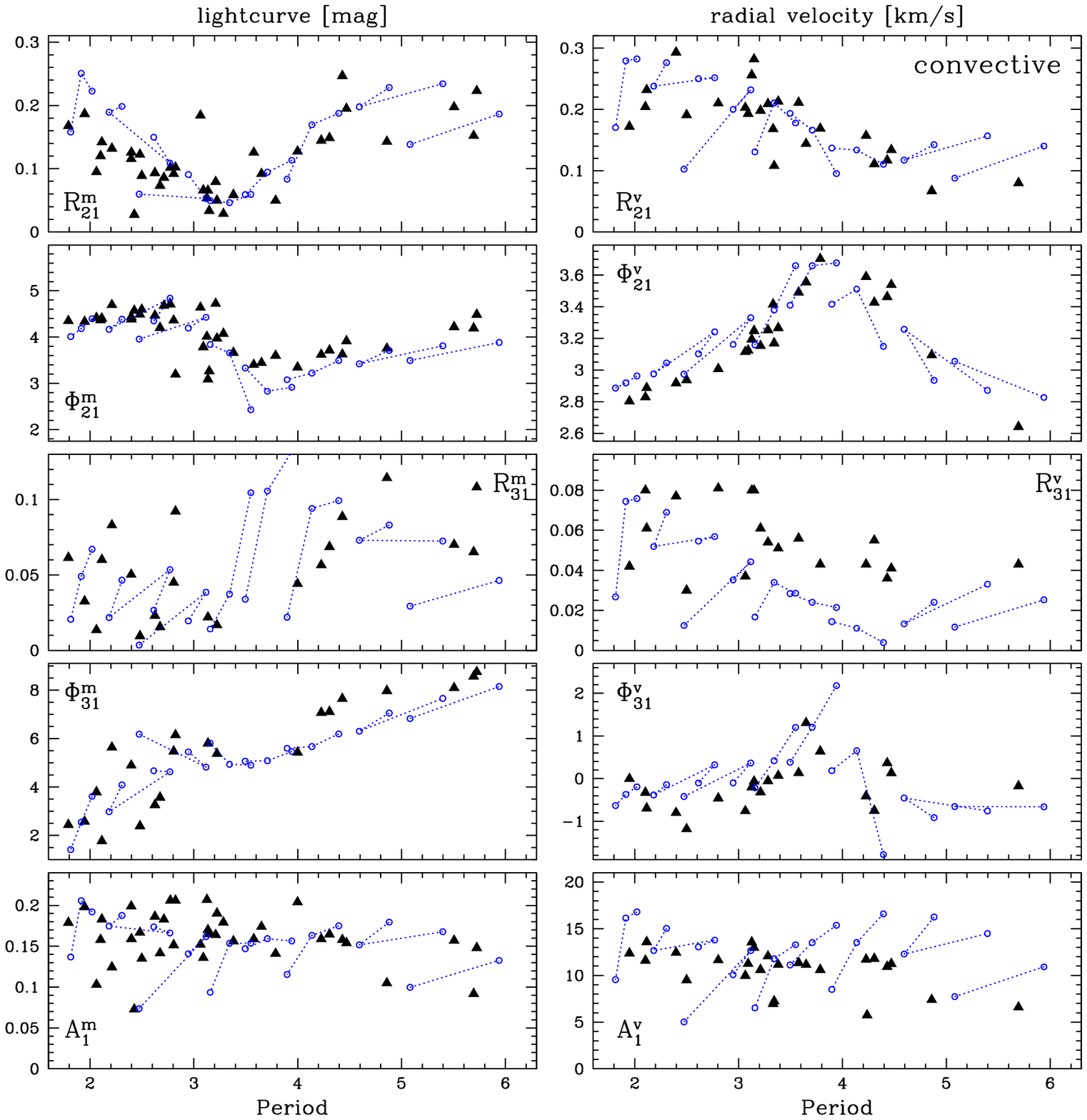}}}

  \noindent{\small 
            Fig.~10: Fourier coefficients $R_{21}$ and $\Phi_{21}$ of convective
            models series \SB\ with the convective enthalpy flux limiter (open
            circles) compared to observations (filled triangles): {\sl Left:}
            light-curve (mag) data; {\sl Right:} Radial velocity (km/s) data.
            The open circles connected by dotted lines refer to sequences with
            the same mass (from left to right: 4.25, 4.50, 4.75, 5.00, 5.15,
            5.25, 5.50, 5.75, 6.00, 6.25 \Mo).  For comparison purposes the
            scales are the same as in Figs. 7 and 12.}

            \vspace{1cm}

\end{figure*}

\vskip 10pt
\centerline{Series \SB}
\vskip 10pt

A striking feature of the convective models of series \SA\ in Section
\ref{s:conv} is the large jump of the $\Phi_{21}^m$, and it is interesting to
see whether the size of this jump can be decreased to observed values by
changing the $\alpha$'s.

First of all it is instructive to investigate whether there are any peculiar
features in the light-curve structure that are connected with that jump.
Fig.~9 (solid line at the top) shows the light-curve of a model of series \SA\
which is located just to the left of that jump.  The light-curve exhibits a
shoulder on the rising branch that is absent in the observed light-curves.
This shoulder appears only in models near the $\Phi_{21}^m$ jump and no
corresponding feature can be found in the radial velocity curve.  If the
pulsation amplitude of the model is increased beyond the observed value through
a decrease in the turbulent viscosity, the shoulder becomes increasingly
pronounced, as the lower solid lines indicate.  Eventually a spike develops
that is similar to the one found in the convective models of RR Lyrae stars
(Feuchtinger 1999b).

In order to cure the problem of the spike Wuchterl \& Feuchtinger (1998) capped
the size of the correlations $\langle s'u' \rangle \approx \langle h'u'
\rangle$ to which both the source of turbulent energy and the convective flux
are proportional (flux limiter).  In series \SB\ we apply the same type of
limiter to the first overtone Cepheid models.  However, in contrast to the RR
Lyrae models we use a higher value of $\alpha_{\rm L}$ = 3 instead of 1 which
diminishes the effect of the flux limiter and hence only slightly changes the
convective structure of the models.  Because the limiter reduces the amount of
convection and therefore also the dissipation, we need to increase the
turbulent viscosity parameter $\alpha_{\mu}$ from 0.25 to 0.33 to maintain the
same pulsation amplitudes.

The resulting change in the light-curve structure can be inferred from the
bottom of Fig.~9 which plots the flux limited light-curve (solid line) as
compared to the nonlimited case (dashed line).  The Fourier analysis yields a
drop of $\Phi_{21}^m$ from 5.42 to 4.20 for the limited model.  The comparison
of the whole flux-limited sequence with observations is given in Fig.~10.  The
bottom panels show $R_{31}$, $\Phi_{31}$ and $A_1$.  It turns out that the
inclusion of the flux limiter decreases the jump in $\Phi_{21}^m$ considerably,
while all other quantities remain almost unaffected.  Clearly the best results
are obtained when a flux limiter is included.

All our attempts to achieve the same effect as obtained with a flux limiter by
using various combinations of $\alpha$'s have proved in vain.  Essentially the
same was found for RR Lyrae stars (Feuchtinger 1999b).  This state of affairs
is somewhat disconcerting because of the {\sl ad hoc} nature of the flux
limiter, and its cause may well be found in the oversimplified nature of our 1D
treatment of turbulent convection.

\vskip 10pt
\centerline{Series \SC}
\vskip 10pt

Another effect that was omitted in the model series \SA\ of
Section~\ref{s:conv} concerns the decrease of turbulent kinetic energy through
radiative losses.  This effect is important when the radiative diffusion time
scale becomes comparable to or smaller than the typical eddy rise time, \ie
when the P\'eclet number is small (This effect is treated differently in the 
Vienna code (Wuchterl \& Feuchtinger 1998) and in the Florida code (Buchler \&
Koll\'ath 2000; Koll\'ath \etal 2000) who follow the recipe of Canuto \& Dubikov
1998).  A nonzero value of the corresponding parameter $\gamma_{r}$ causes both
a decrease of the convective flux and the turbulent kinetic energy.  In our
sequence \SC\ we use $\gamma_{r}$ = 3.5.  To compensate for the resulting decrease
of dissipation and to avoid too large an instability strip and too large
pulsation amplitudes, we increase the mixing length parameter
$\alpha_{\scriptscriptstyle \rm ML}$ from 1.5 to 2 and the turbulent viscosity
$\alpha_{\mu}$ from 0.25 to 0.35 (series \SC, see also Table~1).  This yields
approximately the same pulsation amplitudes as obtained without the P\'eclet 
correction.

 \begin{figure*}
 \vskip 10pt
   \centerline{\vbox{\epsfxsize=18cm\epsfbox{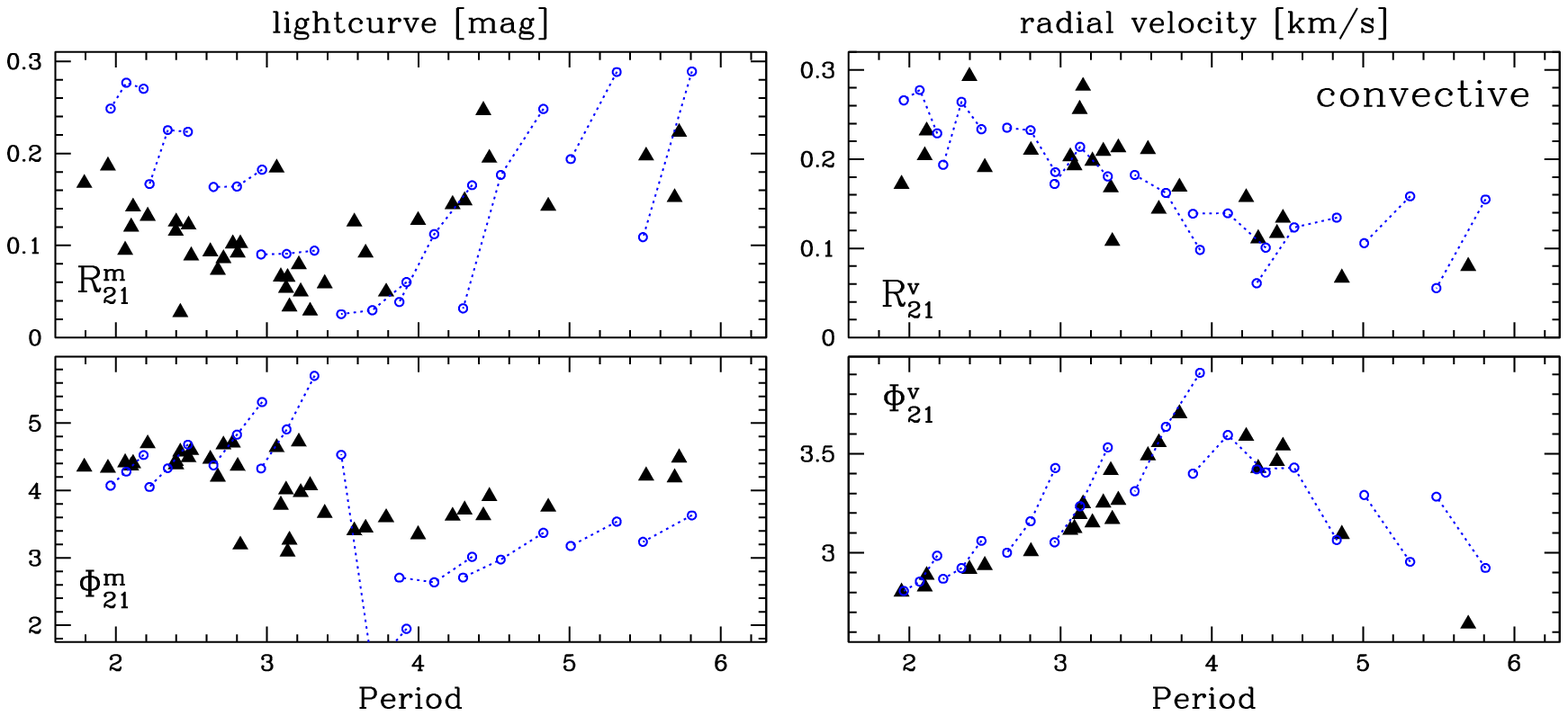}}}

  \noindent{\small Fig.~11: Fourier coefficients $R_{21}$ and $\Phi_{21}$ of
            convective models (series \SC) that include radiative losses (open
            circles) compared to observations (filled triangles): {\sl Left:}
            light-curve (mag) data; {\sl Right:} Radial velocity (km/s) data.
            The open circles connected with by lines refer to sequences with
            the same mass (from left to right: 4.50, 4.75, 5.00, 5.25, 5.50,
            5.75, 6.00, 6.25, 6.50, 7.00 \Mo).  For comparison purposes the
            scales are the same as in Figs. 8 and 10.}

\end{figure*}

The influence on the linear IS boundaries is shown in Fig.~12.  The solid lines
refer to models including radiative losses (\SC), dashed to the original
sequence (\SA), and {\bf F} and {\bf O} denote the fundamental and first
overtone mode, respectively.  Both fundamental and overtone blue edges are
shifted to the blue by the same amount of about 100K.  In contrast, the average
fundamental red edge shift of about 550\K\ to the blue edge is much larger than
the corresponding 200\K\ for the overtone red edge.  Considering the average
linear IS widths (taken at 6 \Mo) we end up with 580\K\ for the first overtone
and 780\K\ for the fundamental, compared to 700\K\ and 1200\K, respectively,
for the series without radiative losses.  Consequently, the inclusion of
radiative losses has a differential effect on fundamental and first overtone
growth rates, which is important for the calibration of the whole Cepheid
picture (\cf~Section \ref{s:fund}).

The nonlinear results for series \SC\ are shown in Fig.~11 and compared to
observed values.  Even though the topology of the IS is changed considerably,
the influence on the Fourier coefficients is not conspicuous.  In particular
the large jump of $\Phi_{21}^m$ is only slightly reduced compared to series
\SA\ in Fig.~6.  Additionally, the position of that jump and also the maximum
of $\Phi_{21}^v$ remain at the same place.  Bearing in mind that several
constraints involving fundamental and double-mode pulsations have not been
considered so far, such insensitivity is welcome because it provides leeway
for matching additional constraints (\cf~Section \ref{s:fund}).

  \vskip 10pt
  \centerline{\vbox{\epsfxsize=9cm\epsfbox{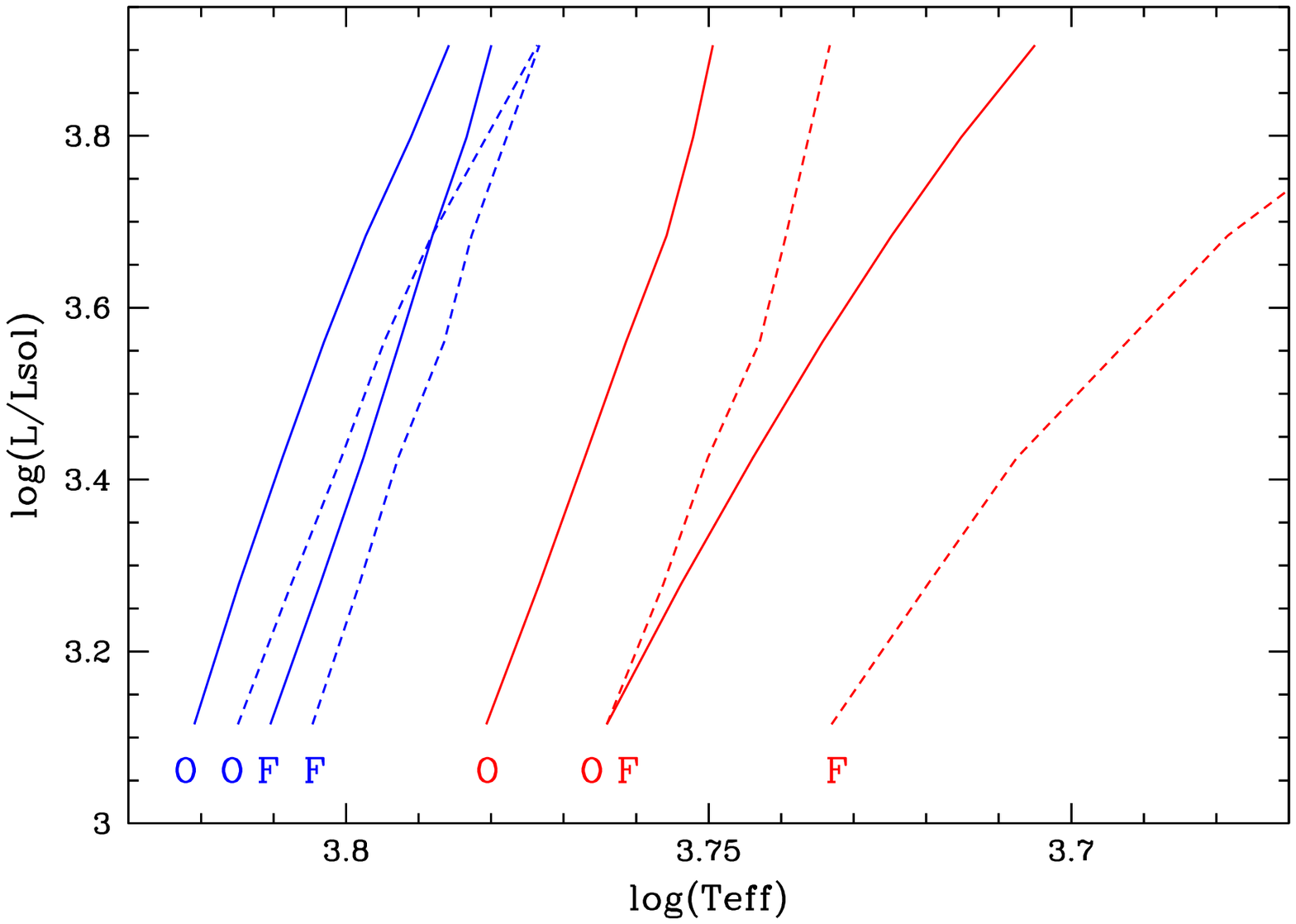}}}

  \noindent{\small Fig.~12: Linear IS boundaries for
           convective model series \SC\ (which include radiative losses, solid
           lines) compared to series \SA\ (dashed lines) in the
           HR diagram.}

  \vspace{1cm}

\vskip 10pt
\centerline{Series \SD\ and \SE}
\vskip 10pt

The Kuhfu\ss~standard choice for $\alpha_{\rm s}$, $\alpha_{\rm c}$ and $c_{\rm
D}$ which gives the mixing length theory (MLT) limit in the local and static
case, leads to rather high values of the turbulent kinetic energy $e_t$.  For a
typical hydrostatic initial model $e_t$ peaks around 0.55$e$ in the H
ionization zone and at 0.25$e$ in the HeI zone, where $e$ denotes the internal
energy.  Dynamical effects might lead to even higher values of $e_t$ during
some stages of the pulsation cycle (\cf Buchler, Yecko, Koll\'ath \& Goupil
1999, Figs. 1 and 2).  The corresponding convective Mach numbers $\sqrt{2/3 \,
e_t} / c_s$, where $c_s$ denotes the adiabatic sound-speed, reach values of
about 0.7.  Clearly one is close to the limit of validity of our convection
model, which, by disregarding pressure fluctuations, assumes a convective
element always to be in pressure equilibrium with its surroundings.  It is
therefore interesting to compute a model series with considerably lower $e_t$.
This can be accomplished in different ways because several $\alpha$ parameters
(\viz $\alpha_{\scriptscriptstyle \rm ML}$, $\alpha_{\rm s}$ and $c_D$) exhibit
a strong influence on $e_t$.

In series \SD\ of Table~1 we increase the dissipation parameter $c_D$ by a
factor of 4, which leads to an average reduction of $e_t$ by a factor of 3.  At
the same time we increase $\alpha_{c}$ to 1.5 times its original value (series
\SA), which results in approximately the same convective flux structure.
Moreover, in order to obtain the right pulsation amplitudes one needs to
increase the turbulent viscosity.  Despite these rather dramatic changes of the
$\alpha$'s only minor changes in the pulsational properties of the models are
found.

Series \SE\ includes both turbulent kinetic energy flux $F_t$ and turbulent
pressure $p_{\rm t}$, but has the same $\alpha$'s as the low $e_t$ series \SD.
The flux $F_t$ has only a small effect on the pulsation for reasonable values
of $\alpha_{\rm t}$, \ie as long as the convection zones do not invade the
outer boundary.  The turbulent pressure is also unimportant as long as it
remains small compared to the gas pressure.  The inclusion of these quantities
thus causes neither significant changes in the topography of the instability
strip nor in the Fourier coefficients.

For some choice of the parameters $\alpha_{\scriptscriptstyle \rm ML}$,
$\alpha_{\rm s}$, $\alpha_{\rm c}$ and $c_{\rm D}$ the turbulent kinetic energy
$e_{\rm t}$ is very large.  Then, because $p_{\rm t} = \alpha_{\rm p} \rho \,
e_{\rm t}$, the turbulent pressure can get as large or even larger than the gas
pressure for the standard value of the parameter $\alpha_{\rm p}$ = 2/3.  Since
a much smaller value of $\alpha_{\rm p}$ does not seem appropriate in this
picture (/eg Baker 1987) this suggests that it would be preferably to use sets
of $\alpha$'s that yield a lower $e_{\rm t}$ profile and a reasonable $p/p_t$
ratio.  On the other hand, such a problem might also reflect the limitations of
the simple 1D model of convection that we use.

\section{Width of the Instability Strip}

We recall that the left (hot) side of the IS determined by the linear
growth-rates (which change sign there), but that the red (cool) edge is
determined by nonlinear effects, namely instability of the limit-cycles.  At
low masses (and luminosities) the overtone limit-cycles become unstable to
double-mode pulsations, and at higher masses they turn into fundamental
pulsations (Udalski \etal 1987, Koll\'ath \etal 2000).

The comparison of our calculated Fourier data with the observations suggests
that the overtone IS must be very narrow.  Indeed, Figs.~6, 10 and 11 show a
strong tendency for the computed values of the $\Phi_{21}^m$ (dotted lines) to
climb above the observed values as the period of the models increases along
each mass sequence, in particular the low mass sequences.  Had we chosen
$\alpha$'s that yield a much broader IS then the disagreement of the computed
values with the observations would have been severe.

It is somewhat puzzling that the observations show practically no low amplitude
overtone Cepheids (Fig.~3), neither in light nor in radial velocity, and
neither at the blue edge nor at the red edge.  Of course there is some
observational bias against low amplitude pulsators but we do not believe that
it can account for the observed deficiency.  In Buchler, Koll\'ath \&
Feuchtinger (2000) we show that the build-up of the pulsation amplitude can be
delayed by stellar evolutionary effects.  But this happens only on the redward
entry into the IS.  Another possibility is that the behavior of the
growth-rates with \Teff\ is much steeper than our calculations indicate.  If
this were the reason it would point to an inadequacy of the simple 1D treatment
of convection that we use.

\section{Fundamental mode pulsators} \label{s:fund}

Even though our first overtone Cepheid models display good agreement with
observations, this tells only one part of the story.  A comprehensive model for
Galactic Cepheids will have to reproduce the observed behavior of the complete
modal behavior (fundamental, overtone and double-mode pulsations) throughout
the whole IS.  Accordingly, further constraints such as the Hertzsprung
progression of the Fourier coefficients of the fundamental Cepheid light- and
radial velocity curves (connected with the P$_0$/P$_2 = 2$ resonance), or the
location and properties of the double-mode pulsations need to be included.
Such a calibration is beyond the scope of this paper. There is no \apriori
guarantee that our adopted parameter sets, which give good results for first
overtone Cepheids, also work for fundamental Cepheids.  We thought it useful to
ascertain that with our $\alpha$'s the fundamental mode models are at least
reasonably good.  On the basis of a few sequences of models we find that even
though the agreement is not perfect, the main features in the Fourier
coefficients can be reproduced.  There is therefore hope that future work will
be able to determine a set of $\alpha$'s that will yield a comprehensive
picture of the Galactic Cepheids.

\vfill\eject

\section{Low metallicity Cepheids}

The Magellanic Clouds are thought to be metal-deficient compared to the Galaxy,
and presumably so are the SMC and LMC Cepheids.  Nevertheless, the observed
characteristics of these Cepheids (\eg stellar parameters, pulsation
amplitudes, position of resonances, double-mode behavior, etc.) are very close
to those of their Galactic siblings.  However, current models show a strong
metallicity (Z) dependence that is in conflict with the observed behavior (\eg
Buchler, Koll\'ath, Beaulieu \& Goupil 1996, Buchler 2000).  This issue will be
addressed in detail in a forthcoming paper.

\section{Summary and conclusions}

In this paper we have addressed the modelling of Galactic first overtone
Cepheids with two different state-of-the-art stellar pulsation codes.  Both
codes include a treatment of time-dependent convective energy transfer, \viz
the Vienna and the Florida codes.  A reexamination of radiative models with an
adaptive mesh and radiation hydrodynamics code reveals no improvement when
compared with the simpler Lagrangean radiative diffusion code.  In particular,
the conspicuous Z-shape of the $\Phi_{21}^m$ with period cannot be reproduced
with radiative modelling.

In contrast, we demonstrate that with the inclusion of convective energy
transport it is possible to reproduce the observed behavior of Galactic first
overtone Cepheids.  The Schaller \etal M--L relation that we have used here
puts both the overtone P$_1$/P$_4$=2 and the fundamental P$_0$/P$_2$=2
resonances in approximately the right places as the agreement between the
calculated and the observed Fourier data show. With a slight adjustment of the
M--L relation the agreement with the observations could be further improved.
In particular our models reveal that the P$_1$/P$_4 =2$ resonance which is
responsible for the structure in the Fourier coefficients, is located at
pulsation periods in the vicinity of $P_1$= 4\dotd 2, as conjectured by Kienzle
\etal (1999) on the basis of their radial velocity data.

\section{Acknowledgements}

This work has been supported by NSF (grant AST 9819608) and by OTKA (T-026031).

\vskip 10pt

{}
\end{document}